\documentstyle[epsf]{aa}

\newcommand{\myfigure}[4]{ 
        \begin{figure*} 
        \setbox100=\hbox{ 
        \epsfxsize=#1 cm 
        \epsfbox{#2.ps}} 
        \centerline{\hbox{\box100}} 
        \caption{#3} 
        \label{#4} 
        \end{figure*} 
} 

\begin{document} 
 
\title {Spectroscopic analysis of two peculiar emission line stars: 
RJHA\,49 \& SS73\,21\thanks{Based on observations made with the 1.52m telescope  
at the European Southern Observatory (La Silla, Chile) under the 
agreement with the Observat\'orio Nacional, Brazil.}}
 
\author{C.B. Pereira\inst{1}, W.L.F. Marcolino\inst{1}, M. Machado\inst{2}
  \& F.X. de Ara\'ujo\inst{1}}
 
\institute{Observat\'orio Nacional-MCT, Rua Jos\'e Cristino, 77. CEP 20921-400.
S\~ao Crist\'ov\~ao. Rio de Janeiro-RJ. Brazil.\\
  \email{claudio@on.br, wagner@on.br, araujo@on.br}
\and
Departamento de F\'{\i}sica, UERJ, Rua S. Francisco Xavier, 400. CEP
12322-100, Rio de Janeiro-RJ. Brazil}
 
\offprints{C.B. Pereira} 
                                
\date{Received . . . ; accepted . .  } 
 
\abstract 
{}
{To investigate the spectra and the evolutionary stages of two peculiar emission-line
stars: RJHA 49 and SS73 21.}
{We used low and high resolution optical data. Line identifications
  and measurements were performed for several features in their spectra.}
{For each object, we have derived the extinction and the excitation
temperature from a set of [Fe\,{\sc ii}] lines, and the electron density from 
[N\,{\sc ii}] lines. For RJHA 49, no detailed spectroscopic study was 
done so far. Regarding SS73 21, our low resolution spectrum have 
confirmed the main characteristics found in previous works. On the other 
side, from our high resolution data, we have found that the H$\alpha$ line
presents a double-peak, in contrast with the suggestion in the literature 
that it should reveal a P-Cygni profile. Surprisingly, we found 
a few He\,{\sc i} transitions resembling P-Cygni profiles (e.g.  He\,{\sc i}
$\lambda 5876$), directly suggesting that mass loss is active in SS73 21. 
We also discussed the nature of these two objects based on the data obtained. 
Although the evolutionary status of SS73 21 seems well established from 
previous studies (a proto-planetary nebula), the situation for RJHA 49 is 
not so clear mainly due to its unknown distance. However, from the strength of 
[N\,{\sc ii}] $\lambda 5754$ relative to [O\,{\sc i}] $\lambda 6300$, 
the possibility of RJHA 49 being a LBV object is reduced, and 
a B[e]-supergiant or a proto-planetary nebula status is more 
plausible.}  
{}
  
 
\titlerunning{Two peculiar emission line stars}

\maketitle
 
\section{Introduction} 

\par The spectral characteristics of a group of stars known as ``B[e]-stars''
(B-type stars with forbidden emission lines, mainly from iron, in the optical
spectrum) have been received wide attention in the recent years. Interestingly, 
different groups of objects in distinct and well defined evolutionary stages 
may present a very similar spectrum, with Balmer lines, several 
permitted and forbidden iron emissions, as well as an infrared excess.
Due to this fact, it was proposed in the literature that all these objects 
should be categorized as stars with the "B[e]-phenomenon" (Lamers et al. 1998).
Yet, there is a group that still resists to be properly identified, namely, 
the ``unclassified B[e] stars'' (or unclB[e] stars). Although some stars 
within this group are better studied than others (e.g. HD 45677 and HD 50138; 
Lamers et al. 1998), the nature and the evolutionary status 
of most of them has not yet been revealed. An interesting   
review regarding the observational properties of different subgroups 
of B[e]-stars and a comparison to other peculiar emission line objects, 
including the unclB[e] class, was recently presented by Miroshnichenko (2006).

\par Following our program to investigate emission-line objects in the 
southern hemisphere at the European Southern Observatory (ESO) whose nature is
not well established, we present in this paper spectroscopic data of two
peculiar emission-line stars: RJHA\,49 (=MWC 819) and SS73\,21 (=Th 35-27).
Both objects were selected from the works of Sanduleak \& Stephenson (1973)
and Allen \& Swings (1976). Previous efforts of our program led to a
classification of 33 emission-line stars (Pereira et al. 2003a) and an
analysis of three stars with $\eta$-Car spectrum: SS73\,11 (Landaberry et al.
2001), SS73\,56, and Hen 2-79 (Pereira et al. 2003b).

\par There are not many objects that have an $\eta$-Car type spectrum.
In the early 70's, Swings and Allen (1973) realized that the spectral
characteristics in the visual of MWC 645 and MWC 819 (=RJHA\,49) looked very
similar to $\eta$-Car. MWC 645 would be later investigated by Jaschek et al.
(1996). RJHA\,49 has already been classified as a possible planetary nebula
(Kohoutek 1971), a Be star with infrared excess (Allen \& Swings 1976) and
more recently, as a B[e] star by The et al. (1994) and an unclB[e] by Lamers
et al. (1998). According to Miroshnichenko (2006), RJHA\,49 is one of the
objects that ``received almost no attention since the introduction of B[e]
stars''. On the other side, the nature of SS73\,21 seems to be better
established. On the basis of infrared {\it IRAS} colors, Parthasarathy \&
Pottasch (1989) first suggested that this object could be a proto-planetary
nebula. Indeed, different recent studies based on images and low resolution
spectroscopy have supported this view (Garc\'{\i}a-Lario et al. 1999; Sahai et
al. 1999; Parthasarathy et al. 2001). In the present paper, the high
resolution optical spectrum of SS73 21 is investigated for the first time.

\par The rest of the paper is divided in the following manner: 
in Sect. 2 we present the details of our observational data, including the
reduction procedure, and the extinction derived for each object. In Sect. 3 we
present some physical conditions derived for both objects, namely, the
excitation temperature from forbidden iron lines and the electron density from
forbidden nitrogen lines.  In Sect.4 we discuss their nature and finally in
the last section we summarize the main points of our work.

\section{The Data} 
 
\subsection{Observations \& Reduction} 
 
\par The low resolution spectroscopic observations were performed using a Boller \& 
Chivens spectrograph at the Cassegrain focus of the ESO 1.52m telescope in La
Silla (Chile) on February 4th, 1999 (RJHA 49) and March 2nd, 1999 (SS73 21).
A UV-flooded thinned Loral Lesser CCD \#39 (2048 x 2048, 15$\mu$m/pixel) was
used as the detector; it gives a high quantum efficiency in the blue and in
the UV. We used the grating \#23 with 600 l/mm, providing a resolution of
about 4.6\AA\, in the range $\sim 4000-8000$\AA. The slit orientation was
East-West and the slit width was 2$''$. The sky conditions in these
observations were mostly clear, but not photometric with a mean seeing of
1.5$''$, therefore the flux calibration should be seen with caution.
 
\par The spectra were reduced using standard IRAF tasks, from bias 
subtraction and flat-field correction, through spectral extraction and
wavelength and flux calibration. Spectrophotometric standards from Oke (1974)
and Hamuy et al. (1994) were observed.

\par In the linearized spectra, the fluxes of emission lines have been 
measured by the conventional method adjusting a gaussian function to the line
profile thereby obtaining the intensity, the central wavelength and the line
width at half power level. Uncertainties in the line intensities come mainly
from the position of the underlying continuum. We estimate the errors in the
fluxes to be about 20\% for weaker lines (line fluxes about 10 on the scale of
H$\beta$=100) and about 10\% for stronger lines. 

\par RJHA 49 and SS73 21 were also observed in high resolution mode 
with FEROS in the 1.52m ESO telescope in La Silla (Chile) on February 9th and
February 3rd of 2001, respectively. The FEROS spectral resolving power is
$R=48\,000$, corresponding to 2.2 pixels of $15\,\mu$m.  The total wavelength
coverage is $\sim 4000-9200$\AA, and the nominal S/N ratio measured by RMS
flux fluctuation is approximately 100 after 3600 secs of exposure time. The
spectra were reduced with the MIDAS pipeline reduction (Kaufer et al. 1999)
package consisting of the following standard steps: CCD bias correction,
flat-fielding, spectrum extraction, wavelength calibration, correction of
barycentric velocity, and spectrum rectification.
 
\subsection{The spectra of RJHA 49 and SS73 21} 
 
\subsubsection{Low resolution} 

\myfigure{16.0}{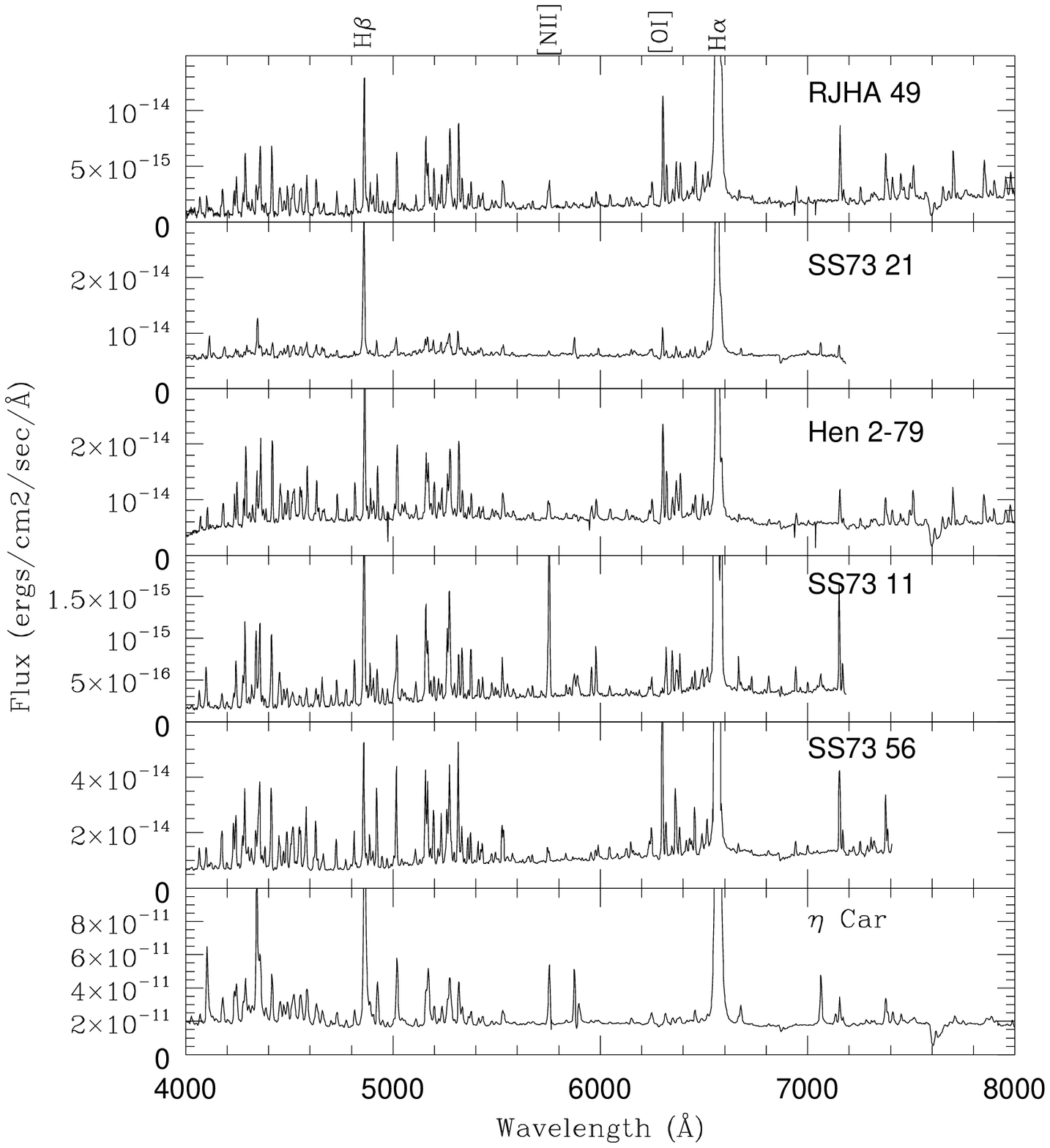}{Optical spectrum of RJHA 49 and SS73 21 in comparison 
with other peculiar emission-line stars already analyzed  
and $\eta$ Car.}  
{fig1} 
 
\par Our optical spectra of RJHA 49 and SS73 21 is displayed in Figure 1.  
For comparison, we also show in this Figure spectra of other objects
previously analyzed by us, namely, Hen 2-79, SS73 11, and SS73 56, as well
as the spectrum of $\eta$ Car, obtained at the same resolution.
 
\par The spectrum of RJHA 49 presents several 
strong emission lines mostly due to single ionized forbidden and permitted
iron over a flat continuum. The majority of the iron features are also present
in the spectrum of the objects above mentioned. The nitrogen forbidden line at
5754\AA\, is present but weaker than the oxygen forbidden line at 6300\AA. The
intensities of these two transitions are particularly important, since they
can be used as a criteria to distinguish for example, a B[e]sg from a LBV star
(Zickgraf 1989). As in the other stars, H$\alpha$ and H$\beta$ are among the
strongest lines in the spectrum.
 
\par G\'arcia-Lario et al. (1999) studied the low resolution spectrum  
of SS73 21 from $3500-11200$\AA. According to these authors, a rich emission
line spectrum can be seen, which is characterized by strong and broad emission
of H\,{\sc i} Balmer lines. Permitted emission lines of He\,{\sc i}, Fe\,{\sc
  ii}, O\,{\sc i} and Ca\,{\sc ii} as well as forbidden transitions from ions
such as [Fe\,{\sc ii}], [N\,{\sc ii}], [O\,{\sc i}], [S\,{\sc ii}] and
[Ca\,{\sc ii}] are also present.  These characteristics are confirmed by our
more recent data.  Furthermore, we found no significant line variations
compared to their study. Nevertheless, as will be seen later in the paper, the 
analysis of the high resolution spectrum of SS73 21 provided valuable 
additional informations compared to the low resolution data.

Overall, we can see from Figure 1 that the spectrum of RJHA 49 and SS73 21 are 
similar to the spectrum of the objects Hen 2-79, SS73 11, SS73 56, and $\eta$
Car. The main characteristics in common are clearly the presence of several 
Fe\,{\sc ii} and [Fe\,{\sc ii}] emissions, and the H\,{\sc i} Balmer lines. 
Interestingly, with the exception of $\eta$ Car, the nature of all these 
peculiar stars is not yet firm established. For SS73 11 for example, although Landaberry et al. 
(2001) could conclude that this object is not a B[e]sg, a HerbigAeB[e] or 
a symbiotic star, these same authors argue that a LBV or a proto-planetary 
nebula classification is possible. For Hen 2-79 and SS73 56, a detailed analysis 
of their spectrum favors a proto-planetary nebula status, but an evolved massive
star classification cannot be discarded  (Pereira et al. 2003b). 
Undoubtedly, one of the main difficulties to 
determine the nature of these and other similar objects is their unknown 
distance. This point and the status of RJHA 49 and SS73 21 will be discussed 
in Section 4.

\par In Table 1 we provide the line identifications as well as the line fluxes 
of transitions other than from Fe\,{\sc ii}, for RJHA 49 and SS73 21. 
As it can be seen, some features in the spectrum of SS73 21 are not present 
in RJHA 49, such as the He\,{\sc i} and the [S\,{\sc ii}] lines. For RJHA 49, 
there are no previous line measurements in the literature. In addition to
Table 1, columns 4th and 7th of Table 2 lists the line flux, in units of 
H$\beta$=100, of some multiplets of single ionized forbidden iron used for 
reddening and excitation temperature determination (see Sections
2.3 and 3.1). For line identifications we 
used the same procedure described in Landaberry et al.(2001).

\begin{table} 
\caption{Observed emission lines other than Fe\,{\sc ii} identified in the 
spectrum of RJHA 49 and SS73 21 in units of H$\beta$=100.0.}  
\begin{tabular}{|l|c|c|c|}\hline 
Wavelength & Identification & \multicolumn{2}{c|}{F($\lambda$)}\\\hline 
& & RJHA 49 & SS73 21 \\\hline 
4101 & H$\delta$ & 11.1 & 14.0 \\ 
4340 & H$\gamma$ & 25.7 & 23.8 \\ 
4861 & H$\beta$  & 100.0 & 100.0 \\ 
5754 & [N\,{\sc ii}] & 18.3 & 3.5 \\ 
5876 & He\,{\sc i} & ---  & 10.5 \\ 
6300 & [O\,{\sc i}] & 82.4 & 16.2 \\ 
6363 & [O\,{\sc i}] & 33.1 &  8.6 \\ 
6563 & H$\alpha$ & 5110.0 & 855.0 \\ 
6584 & [N\,{\sc ii}] & 100.0 & 43.1 \\ 
6678 & He\,{\sc i} & --- & 3.0 \\ 
6717 & [S\,{\sc ii}] & --- & 1.3 \\ 
6731 & [S\,{\sc ii}] & --- & 1.7 \\ 
7065 & He\,{\sc i} & --- & 7.7 \\\hline 
\end{tabular} 
\end{table} 
 
\begin{table*} 
\begin{center} 
\caption{Multiplets, wavelengths, excitation potential ($\chi$), observed  
emission line fluxes F($\lambda$) in units of H$\beta$=100.0, log $I$  
(defined in the text) and the parameter $\beta$ (also defined in the text)  
of some selected [Fe\,{\sc ii}] emission lines used for reddening and  
excitation temperature determination of  
RJHA 49 and SS73 21.} 
\begin{tabular}{|l|c|c||c|c|c||c|c|c|}\hline 
   & & & \multicolumn{3}{c||}{RJHA 49} & \multicolumn{3}{c|}{SS73 21} \\\hline 
M  & Identification & $\chi$ (eV) & F($\lambda$) & log $I$ & $\beta$ 
&F($\lambda$)  & log $I$   & $\beta$ \\\hline  
4F & 4639.68 & 2.77 & 5.0  &  0.23 & 4.39 & --- & ---   & ---  \\  
   & 4728.07 & 2.12 & 10.0 & -0.12 & 4.41 & --- & ----  & ---  \\ 
   & 4889.63 & 2.57 & 10.0 &  0.10 & 4.38 & 2.5 & -0.50 & 3.77 \\\hline 
6F & 4416.27 & 2.79 & 13.0 & -0.01 & 4.33 & 2.9 & -0.69 & 3.59 \\  
   & 4457.95 & 2.82 & ---  & ---   & ---  & 1.4 & -0.59 & 3.58 \\\hline 
7F & 4287.40 & 2.88 & 20.0 & 0.02  & 4.46 & 5.8 & -0.52 & 3.71 \\  
   & 4359.34 & 2.88 & 16.0 & 0.07  & 4.38 & 4.5 & -0.48 & 3.75 \\  
   & 4413.78 & 2.88 & 13.0 & 0.12  & 4.44 & --- & ---   & ---  \\  
   & 4452.11 & 2.88 &  9.0 & 0.17  & 4.49 & 1.7 & -0.55 & 3.68 \\  
   & 4474.91 & 2.88 &  2.0 &-0.18  & 4.14 & --- & ---   & ---  \\\hline 
14F& 7155.14 & 1.96 & 54.0 & 0.81  & 4.71 & 8.1 & -0.01 & 4.14 \\  
   & 7171.98 & 2.01 & ---  & ---   & ---  & 2.2 & -0.02 & 4.10 \\  
   & 7388.16 & 2.02 &  9.0 & 0.72  & 4.59 & 2.2 & -0.05 & 4.24 \\  
   & 7452.50 & 2.02 & ---  & ---   & ---  & 2.6 &  0.04 & 4.16 \\\hline 
17F& 5412.64 & 2.63 & 7.0  & 0.32  & 4.50 & 1.0 & -0.53 & 3.67 \\  
   & 5527.30 & 2.53 & 13.0 & 0.28  & 4.60 & 2.2 & -0.49 & 3.85 \\\hline  
18F& 5107.96 & 2.77 &  3.0 & 0.36  & 4.29 & --- & ---   & ---  \\  
   & 5158.00 & 2.69 & 10.0 & 0.44  & 4.42 & 1.4 & -0.41 & 3.65 \\  
   & 5181.97 & 2.77 &  4.0 & 0.34  & 4.27 & --- & ---   & ---  \\  
   & 5268.88 & 2.69 &  4.0 & 0.25  & 4.23 & --- & ---   & ---  \\  
   & 5273.38 & 2.57 &  --- & ---   & ---  & 3.6 & -0.32 & 3.80 \\  
   & 5433.15 & 2.67 &  8.0 & 0.44  & 4.61 & --- & ---   & ---  \\\hline 
19F& 5111.63 & 2.65 &  4.0 & 0.02  & 4.09 & --- & ---   & ---  \\  
   & 5158.81 & 2.62 & 20.0 &-0.01  & 4.08 & 6.0 & -0.53 & 3.61 \\  
   & 5220.06 & 2.66 &  4.0 & 0.07  & 4.14 & 1.2 & -0.45 & 3.69 \\  
   & 5261.61 & 2.65 & 17.0 & 0.17  & 4.25 & 4.0 & -0.46 & 3.67 \\  
   & 5333.65 & 2.66 &  --- & ---   & ---  & 2.9 & -0.43 & 3.69 \\  
   & 5376.47 & 2.68 & 15.0 & 0.37  & 4.45 & 2.5 & -0.39 & 3.72 \\\hline  
20F& 4814.55 & 2.79 & 9.0  &-0.10  & 3.97 & 2.0 & -0.76 & 3.34 \\  
   & 4874.49 & 2.83 & 2.0  &-0.13  & 3.92 & --- & ---   & ---  \\ 
   & 4905.35 & 2.82 & 9.0  & 0.27  & 4.33 & --- & ---   & ---  \\  
   & 4947.38 & 2.79 & 2.0  & 0.16  & 4.24 & --- & ---   & ---  \\  
   & 4950.74 & 2.84 & 4.0  & 0.36  & 4.41 & 0.5 & -0.55 & 3.52 \\  
   & 4973.39 & 2.83 & 5.0  & 0.36  & 4.41 & 0.5 & -0.64 & 3.43 \\  
   & 5005.52 & 2.82 & 5.0  & 0.52  & 4.58 & --- & ---   & ---  \\  
   & 5020.24 & 2.84 & 3.0  & 0.21  & 4.26 & --- & ---   & ---  \\  
   & 5043.53 & 2.83 & 5.0  & 0.70  & 4.75 & --- & ---   & ---  \\\hline  
\end{tabular} 
\end{center} 
\end{table*} 
 
\begin{table*} 
\begin{center} 
\noindent {\bf Table 2} \\ Multiplets, wavelengths, excitation potential
($\chi$), observed  emission line fluxes F($\lambda$) in units of
H$\beta$=100.0, log $I$  (defined in the text) and the parameter $\beta$ 
(also defined in the text) of some selected emission lines used for reddening 
and excitation temperature  determination of RJHA 49 and SS73 21.
\begin{tabular}{|l|c|c||c|c|c||c|c|c|}\hline 
   & & & \multicolumn{3}{c||}{RJHA 49} & \multicolumn{3}{c|}{SS73 21} \\\hline 
M  & Identification & $\chi$ (eV) & F($\lambda$) & log $I$ & $\beta$ 
&F($\lambda$) & log $I$    & $\beta$ \\\hline  
21F& 4243.98 & 3.14 & 22.0 & 0.00 & 4.17 & 3.3 & -0.83 & 3.26 \\  
   & 4244.81 & 3.21 &  4.0 & 0.03 & 4.16 & 0.9 & -0.62 & 3.43 \\  
   & 4276.83 & 3.19 &  6.0 &-0.32 & 3.83 & 2.2 & -0.75 & 3.16 \\  
   & 4319.62 & 3.21 &  4.0 &-0.29 & 3.84 & 1.4 & -0.74 & 3.30 \\  
   & 4346.85 & 3.14 &  6.0 & 0.08 & 4.25 & 0.8 & -0.80 & 3.29 \\  
   & 4352.78 & 3.19 & 11.0 & 0.29 & 4.42 & 1.7 & -0.47 & 3.52 \\  
   & 4358.37 & 3.22 & 11.0 & 0.14 & 4.27 & 1.9 & -0.26 & 3.42 \\  
   & 4372.43 & 3.21 &  3.0 &-0.13 & 4.00 & 1.2 & -0.53 & 3.52 \\\hline 
31F& 7047.99 & 2.84 & 2.0  & 1.26 & 4.67 & --- & ---   & --- \\\hline  
34F& 5477.25 & 3.32 & 2.0  & 0.18 & 3.83 & --- & ---   & --- \\  
   & 5746.96 & 3.18 & 6.0  & 0.45 & 4.10 & 1.0 & -0.33 & 3.92 \\\hline   
35F& 5163.94 & 3.37 & 8.0  & 0.40 & 4.07 & 1.0 & -0.51 & 3.22 \\\  
   & 5283.10 & 3.37 & 3.0  & 0.29 & 4.01 & --- & ---   & --- \\\hline  
39F& 5551.53 & 3.89 & 3.0  & 0.68 & 3.94 & --- & ---   & --- \\\hline  
43F& 6944.91 & 3.80 & 3.0  & 0.90 & 3.80 & --- & ---   & --- \\\hline  
44F& 6188.55 & 3.95 & 1.0  & 0.94 & 3.87 & --- & ---   & --- \\\hline  
a2G-a2I & 5870.00 & 4.10 & 2.0 & 0.37 & 3.39 & --- & --- & --- \\ 
        & 6044.10 & 4.08 & 3.0 & 0.73 & 3.75 & --- & --- & --- \\\hline 
a2G-b2D & 4898.61 & 4.50 & 4.0 & 0.42 & 3.60 & --- & --- & --- \\  
        & 5060.08 & 4.74 & 3.0 & 0.60 & 3.78 & --- & --- & --- \\\hline  
a2G-c2G & 5673.22 & 4.10 & 6.0 & 0.65 & 3.67 & --- & --- & --- \\  
        & 5835.44 & 4.70 & 5.0 & 0.65 & 3.66 & --- & --- & --- \\\hline  
a2P-c2D & 5048.18 & 4.70 & 3.0 & 0.72 & 3.72 & --- & --- & --- \\\hline 
\end{tabular} 
\end{center} 
\end{table*} 
 
\subsubsection{High Resolution} 
\label{feros}

\myfigure{20.0}{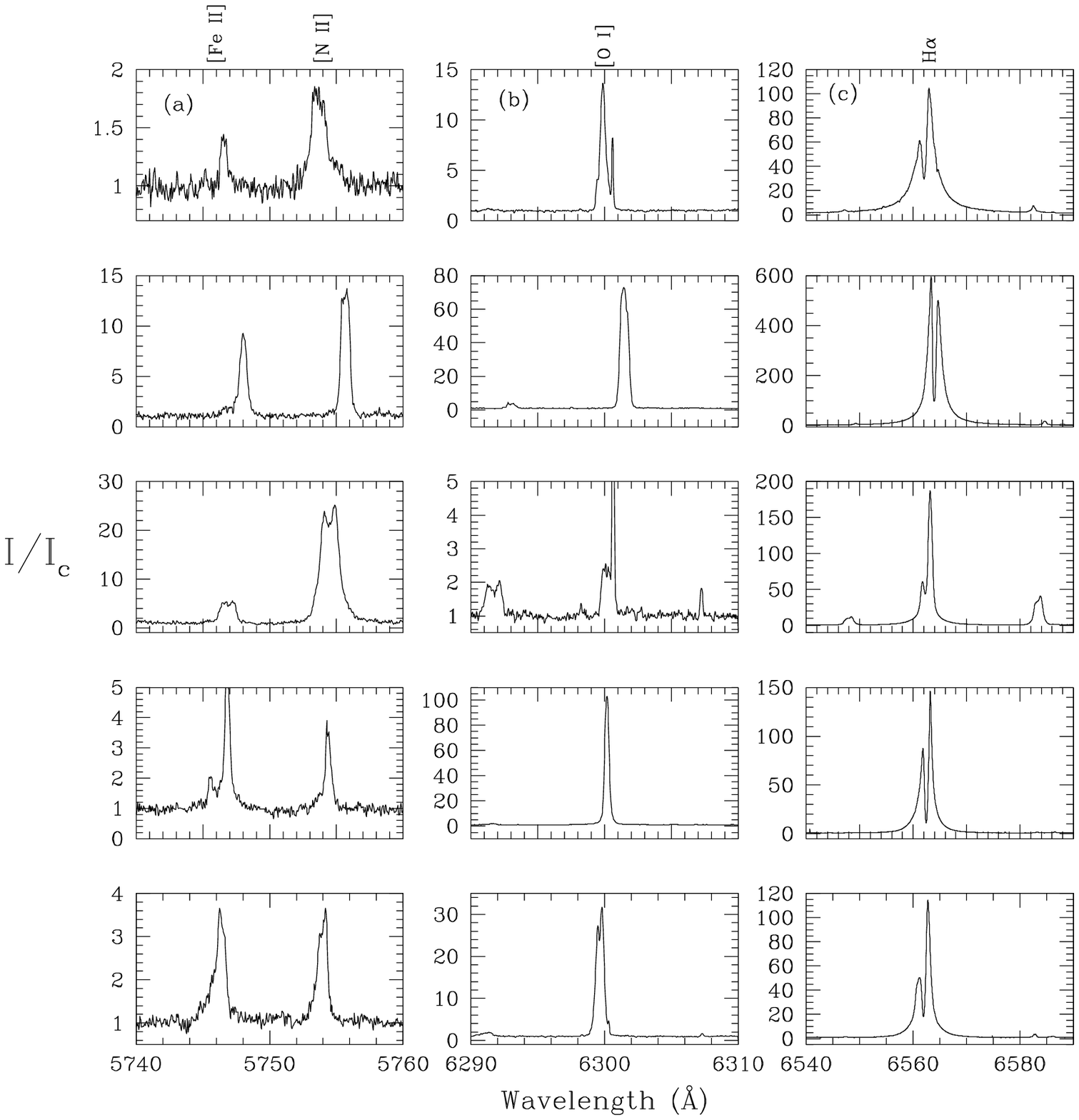}{High resolution spectra obtained with FEROS   
spectrograph in the region around the forbidden nitrogen line at 5754\AA (a), 
6300\AA (b) and H$\alpha$ (c). From top to bottom we show the spectra of 
SS73 21, RJHA 49, SS73\,11, SS73\,56 and Hen 2-79. Notice the strength of 
H$\alpha$ in RJHA 49. The intensities are in continuum units.}
{f2} 

\par The high-resolution spectra were used to better identify some features  
which are blended or were not resolved in the low resolution spectra. This
procedure allowed us to estimate the different contributions of each
transition and discover important line profiles (e.g. H$\alpha$; see below).

\par In Figure 2 we show three spectral regions to better illustrate 
the differences in strength of some emission lines among the objects studied
here (RJHA 49 and SS73 21) to those already studied in Landaberry et al.
(2001) and Pereira et al. (2003b). It can be seen that the strength of the
forbidden line [O\,{\sc i}]$\lambda$6300 is stronger than [N\,{\sc
  ii}]$\lambda$5754 in RJHA 49, SS73 21, SS73 56 and Hen 2-79 while in SS73 11
the opposite occurs.

\par Our optical high resolution spectrum of SS73 21 reveals interesting additional 
informations regarding previous studies. The most important one is that the
H$\alpha$ line clearly presents a double-peak rather than the P-Cygni profile
proposed by Parthasarathy et al. (2001) on the basis of low resolution data.
As can be seen in Figure 2, the H$\alpha$ double-peak profile is also present
in the other stars. SS73\,21 has the broadest profile.  In RJHA\,49, H$\alpha$
has the strongest intensity and the blue peak is more intense than the red
one. Table 3 shows the intensities relative to the continuum and
equivalent widths of some identified lines in the spectrum of these stars.
The 'blue' and 'red' component mentioned in Table 3 refers to the two peaks
seen in the profile of the Balmer lines.

\begin{table*} 
\caption{Equivalent widths and intensities relative to the continuum of some
  lines in the spectra of RJHA\,49 and SS73\,21.}  
\begin{tabular}{|l|c|c|c|c|c|}\hline 
Wavelength & Identification & \multicolumn{2}{c|}{RJHA\,49} 
& \multicolumn{2}{c|}{SS73\,21} \\\hline 
Line       & $\lambda$$_{\rm lab}$ & I/I$_{c}$ & W$_{\lambda}$(\AA) & I/I$_{c}$ & 
W$_{\lambda}$(\AA)\\\hline
H$\gamma$-blue & 4340 & 11 & 3 & 3 & 4\\
H$\gamma$-red  &      & 11 & 6 & 8 & 9\\

H$\beta$-blue  & 4861 & 18 & 11 & 9  & 8\\
H$\beta$-red   &      & 19 & 16 & 23 & 26\\ 

Fe\,{\sc ii}   & 4923 & 23 & 11 & 4 & 2\\

Fe\,{\sc ii}   & 5018 & 36 & 20 & 6 & 3\\

Fe\,{\sc ii}   & 5169 & 23 & 15 & 4 & 2\\

Fe\,{\sc ii}   & 5316 & 45 & 33 & 9 & 5\\

[N\,{\sc ii}]  & 5754 & 12 & 8 & 2 & 1\\

He\,{\sc i}    & 5876 & --- & --- & 4 & 5\\ 

[O\,{\sc i}]   & 6300 & 75 & 38 & 12 & 7\\

H$\alpha$-blue & 6563 & 595 & 77 & 66  & 52\\  
H$\alpha$-red  &      & 502 & 79 & 114 & 82\\

[Fe\,{\sc ii}] & 7155 & 31 & 24 & 5 & 3 \\

O\,{\sc i}     & 8446 & 126 & 150 & 28 & 42\\

Ca\,{\sc ii}   & 8498 & 9 & 7 & 11 & 15\\

Ca\,{\sc ii}   & 8662 & 3 & 2 & 11 & 13\\\hline
\end{tabular}
\end{table*}

\subsection{Extinction}
 
\par We determined the extinction parameter for RJHA 49 and SS73 21 
in the same way as Pagel (1969). Our previous works regarding SS73 11, SS73 56
and Hen 2-79 (Landaberry et al 2001; Pereira et al 2003b) followed the same
procedure. We first measured line fluxes of some Fe\,{\sc ii} forbidden lines
(between 4100\AA\, and 7000\AA) with excitation potential between 2.5 and 3.2
eV. We then plot $log\,I$ (defined below) against the reciprocal wavelength
(1/$\lambda$($\mu$\,m)) in the abscissa. The ordinate $log\,I$ is related to
the difference between the logarithm of the observed flux and the logarithm of
the emitted flux by the source in the same wavelength range according to:
 
\begin{equation} 
log\,I=log\,(F_{\rm obs}(\lambda))-(log\,(5000/\lambda)+log\,(gA)-0.56\,\chi+2.0). 
\end{equation} 
 
\par In the above expression, F$_{\rm obs}$($\lambda$) is the observed line flux in 
units of H$\beta$=100, gA is the statistical weight multiplied by the
transition probabilities, and $\chi$ is excitation potential of the upper
level of the transition. The factor 2.0 accounts for $log$ F(H$\beta$)=100.
 
\par The 5th and 8th column of Table 2 give $log\,I$ as  
defined in the text for both stars.  The color excess E(B-V) which results
from this procedure is 1.22$\pm$0.13 for RJHA 49 and 0.77$\pm$0.07 for SS73
21. For SS73 21, previous extinction determinations are given by
Garc\'{\i}a-Lario et al. (1999) and Parthasarathy et al. (2001). They have
obtained, respectively, 1.3 and 0.75. For RJHA 49 there was no previous
extinction determination in the direction of this object. Probably the main
reason for a difference between our derived value for SS73 21 and the one
obtained by Garc\'{\i}a-Lario et al. (1999) is that the H$\alpha$ shows a double-peak 
profile. Therefore, Balmer lines tend to deviate from pure case B
recombination and due to that, the derereddened H$\alpha$/H$\beta$ ratio 
clearly disagree from case B recombination. A similar effect is seen in 
symbiotic stars (Oliversen \& Anderson, 1982).

\section{Physical Conditions} 
 
\subsection{Excitation temperature} 
\label{temperature}
 
\par Since we observe several emission lines of forbidden single ionized 
iron in the spectra of RJHA 49 and SS73 21, it is possible to derive the 
excitation temperature in the emitting region by following the  
description of Viotti (1969) (see also Thackeray 1967). This procedure  
was previously used by us for SS73 11, SS73 56 and Hen 2-79.  
 
\par In Table 2 we show the parameter $\beta$ (6th and 9th columns) defined as  
$\beta$\,=\,log (F$_{\rm c}$($\lambda$)\,$\lambda$(\AA)/gA) and the excitation
potential of the forbidden lines used in the calculation (3rd column).  In the
above expression F$_{\rm c}$($\lambda$) is the line intensity in units of
H$\beta$=100 corrected for reddening and $\lambda$ is given in Angstroms. The
obtained temperature based on this method is T$_{\rm exc}$=(8\,300$\pm$850)K
for SS73 21 and T$_{\rm exc}$=(12\,000$\pm$1\,100)K for RJHA 49.
 
\subsection{Electron Density} 
 
\par The electron density was obtained using the dereddened [N\,{\sc ii}] 
$\lambda$6584/5754 ratio. For RJHA 49 and SS73 21 the ratios are  
3.0 and 8.3, respectively. Adopting an electron temperature of 10\,000\,K, the 
electron density is 8.2$\times$10$^{5}$ cm$^{-3}$ for RJHA 49 and  
2.6$\times$10$^{5}$ cm$^{-3}$ for SS73 21. Garc\'{\i}a-Lario et al. (1999)  
found a value for SS73 21 compatible to ours ($\sim 2\times 10^{5}$ cm$^{-3}$). 
 
\section{The Nature of RJHA 49 and SS73 21} 

\myfigure{12.0}{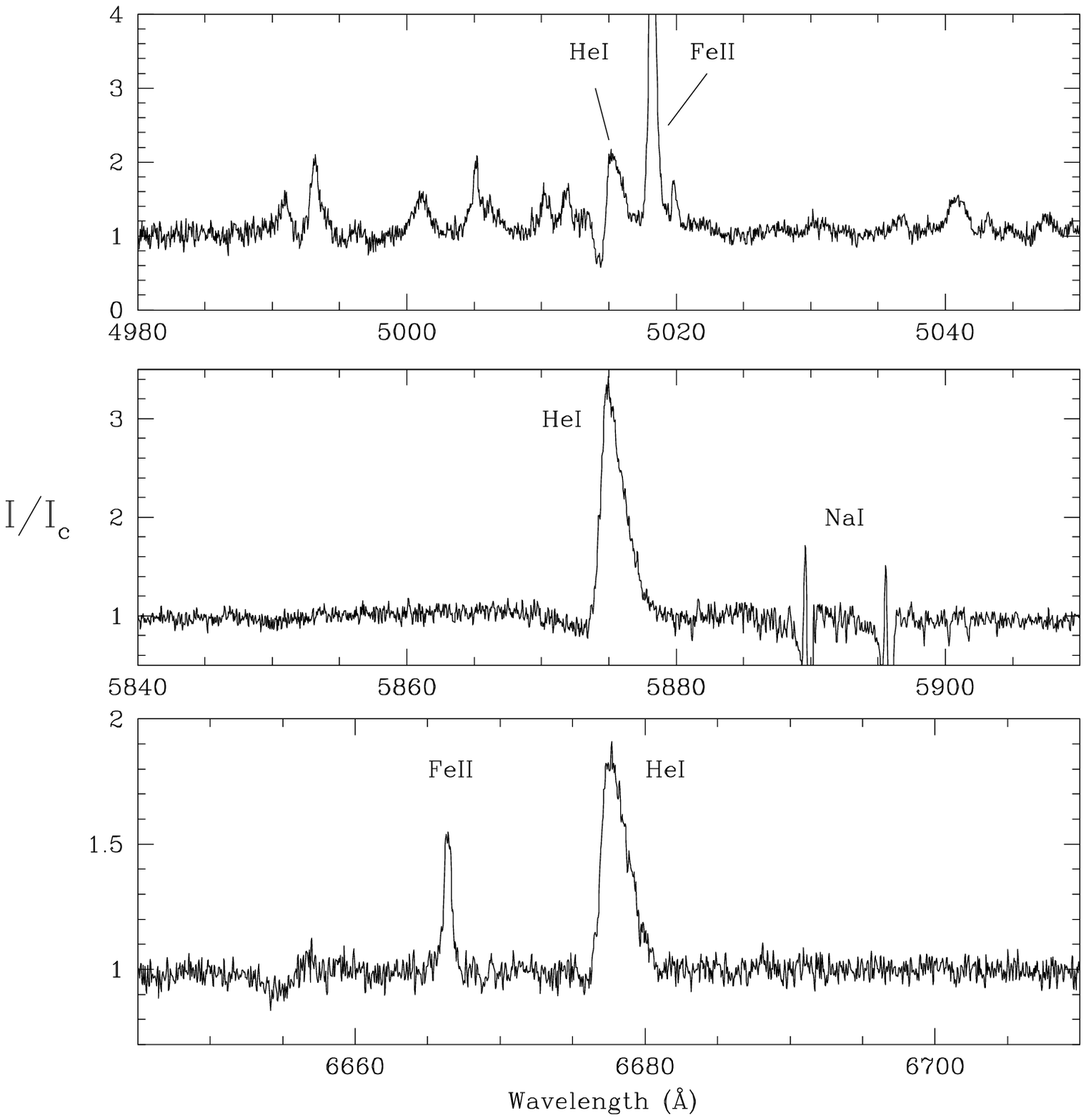}{Helium lines in SS73\,21 with apparent P-Cygni profiles.
The intensities are in continuum units.}  
{f3} 
 
\par Although we have investigated in some detail the spectrum of 
RJHA 49 and have estimated some physical conditions of its envelope, it is
still difficult to determine its evolutionary status since its distance is not
well constrained and thus, its luminosity cannot be accurately derived.  The
lack of wide and narrow band images such as the ones made in the case of SS73
21 (Sahai et al. 1999) also complicates this kind of discussion.  However,
according to Zickgraf (1989), it is possible to use [N\,{\sc ii}] and [O\,{\sc
  i}] to distinguish a B[e]sg star from a LBV. By following this criteria,
since the [O\,{\sc i}] $\lambda 6300$ line is more intense than the [N\,{\sc
  ii}] $\lambda 5754$ (see Figure 2), we conclude that RJHA\,49 is either a
B[e]-supergiant or a proto-planetary nebula.

\par Although the emission-line 
spectra of RJHA 49 and SS73 21 and the other stars look similar to $\eta$ Car
(Figure 1) this does not mean that they are in the same evolutionary stage.
Hillier et al. (2001) reported that the spectrum of $\eta$ Car taken at high
spatial resolution ($0."1 \times 0."13$) with the {\it Hubble Space Telescope}
({\it HST}) is considerably different than the ones obtained from ground
telescopes. Low resolution spectroscopic observation of $\eta$ Car, as the one
presented in our Figure 1, results in a combination of nebular (mainly from
the Weigelt blobs) and central source spectra, i.e., in narrow permitted and
forbidden lines (e.g.  [Fe\,{\sc ii}], Fe\,{\sc ii} and H\,{\sc i})
superimposed on a broad emission line spectrum.

\par As we already mentioned, SS73 21 was first suggested to be a 
proto planetary-nebula by Parthasarathy \& Pottasch (1989), according to its
far-infrared {\it IRAS} colors. Recently, {\it HST} images have revealed long
cylindrical-shaped bipolar lobes, surrounded by a faint elliptical halo, which
is possibly a remnant of the AGB phase (Sahai et al. 1999). These
characteristics were almost simultaneously observed by G\'arcia-Lario et al.
(1999) through images and optical and near-infrared spectroscopy. The presence
of a circumstellar disk is also inferred from these studies and from
polarization data (Scarrot \& Scarrot 1995). 

The amount of reddening in the direction of SS73 21 can help to 
constrain its luminosity and thus its nature. However, different 
values exist for this quantity. If we consider a visual extinction of about 
7 magnitudes, as in the work of Sahai et al. (1999), we would have a 
luminosity of approximately 28000$L_{\odot}$, at a distance of 3kpc 
(Bujarrabal \& Bachiller 1991). This places this star in the HR diagram 
among Be supergiants (Figure 8 of Miroshnichenko et al. 2001), for a 
probable temperature of $\sim 22000$K (Sahai et al. 1999). On the other 
hand, by considering a visual extinction of 2.4 magnitudes (from the present 
work), we arrive at a much lower value for the luminosity, of 2000$L_\odot$. 
This estimate shows a better agreement with a proto-planetary nebula status 
for SS73 21, which is claimed in the literature.

\par Our analysis of the high resolution data of SS73 21 shows 
important spectral characteristics not seen in previous studies. An important
matter which is highlighted in the literature is that the broad, non-gaussian
profile of the H$\alpha$ line seen in low resolution spectra of SS73 21 and of
some other proto-planetary nebula (e.g. M 2-9) is due to mass outflows.
Indeed, Parthasarathy et al. (2001) suggested that the asymmetry in H$\alpha$
seen in SS73 21 is probably due to a P-Cygni profile.  However, as we have
presented in Section 2.2.2, our high-resolution spectrum of SS73 21
reveals a H$\alpha$ with a double-peak profile.  This fact is compatible with
the presence of a circumstellar disk, which was deduced from the studies
aforementioned. 

\par Another very interesting characteristic revealed by the high-resolution 
spectrum of SS73 21 is that some He\,{\sc i} lines resemble a P-Cygni
profile. This can be seen in Figure 3 for the transitions He\,{\sc i} $\lambda
5015$, $\lambda 5876$ and $\lambda 6678$. If these lines are indeed in
P-Cygni, it is a direct evidence that mass-loss is active in SS73 21. We found
no other lines of other ions with such profiles. At present, it would be 
very difficult to determine the formation region of these emissions. A stellar wind
of the central star could be a possibility, but P-Cygni lines can be also
formed in a strong bipolar outflow. Clearly, it would be valuable 
to obtain a mass-loss rate estimate. However, this is beyond 
the scope of the present paper, since it would require the use 
of radiative transfer models. 

An estimate for the expansion velocity involved can be made from 
the He\,{\sc i} $\lambda 5015$ line. We chose not to consider 
the He\,{\sc i} $\lambda 5876$ and He\,{\sc i} $\lambda 6678$ 
lines shown in Figure 3 in this calculation. For the former, 
the wavelength coverage of the absorption part of the profile 
is very uncertain. In the latter transition, the absorption is 
barely visible. Before computing the expansion velocity, we have derived 
a radial velocity ($V_{rad}$) for SS73 21 of approximately +63km s$^{-1}$ 
from a comparison between laboratory and observed wavelengths of several 
emission lines. After correcting for $V_{rad}$, and considering the 
wavelength where the absorption part of the He\,{\sc i} $\lambda 5015$ 
profile returns to the continuum, we finally derived a value of $\sim 90$ km s$^{-1}$. 
This value is lower than the terminal velocities usually found in central 
stars of planetary nebulae, which reach hundreds or even thousands of km
s$^{-1}$ (e.g. Hultzsch et al. 2007), and it is considerably higher than 
the expansion velocity related to molecular transitions of CO in SS73 21, 
which is $\sim 15$ km s$^{-1}$ (Bujarrabal \& Bachiller 1991).

\section{Summary and Conclusions} 

\par We have analyzed low and high resolution optical spectra of 
two peculiar emission-line stars: RJHA 49 and SS73 21. We have 
performed line measurements and identifications for several 
features in their spectra. The spectrum of RJHA 49 was analyzed in 
detail for the first time. For both objects, a set of iron and nitrogen 
forbidden lines were used to derived the interstellar reddening, 
the excitation temperature, and the electron density. We showed that 
the H$\alpha$ line in SS73 21 presents a double-peak profile, in 
contrast with the suggestion in the literature that it should 
reveal a P-Cygni with high resolution data. The presence of 
the double-peak implies that the determination of E(B-V) from 
the H$\alpha$/H$\beta$ ratio is not reliable, since we cannot 
securely assume case B recombination. We found that some He\,{\sc i} lines 
in SS73 21 resemble a P-Cygni profile, directly suggesting that mass-loss 
is active in this star. Finally, we discussed the nature of both objects, where the 
status of SS73\,21 is more clear (a proto-planetary nebula) than for 
RJHA 49. Solely from the [O\,{\sc i}] and [N\,{\sc ii}] line strengths, 
RJHA 49 is either a B[e]-supergiant or a proto-planetary nebula. 
In order to gain more insight on the evolutionary stages of these and 
other similar peculiar emission line stars, distance estimates, high 
angular resolution spectroscopy, as well as wide and narrow band images 
would be very desirable.
 
\begin{acknowledgements} 

W. M. acknowledges CNPq for the financial support (post-doc position - 151635).

\end{acknowledgements} 
 
{} 
 
\end{document}